\documentclass[english,twocolumn,pra,showpacs,longbibliography]{revtex4-1}
\usepackage[T1]{fontenc}
\usepackage[latin9]{inputenc}
\usepackage{verbatim}
\usepackage{bm}
\usepackage{amsmath}
\usepackage{amssymb}
\usepackage{graphicx}
\usepackage{esint}

\makeatletter

\@ifundefined{textcolor}{}
{%
\definecolor{BLACK}{gray}{0}
\definecolor{WHITE}{gray}{1}
\definecolor{RED}{rgb}{1,0,0}
\definecolor{GREEN}{rgb}{0,1,0}
\definecolor{BLUE}{rgb}{0,0,1}
\definecolor{CYAN}{cmyk}{1,0,0,0}
\definecolor{MAGENTA}{cmyk}{0,1,0,0}
\definecolor{YELLOW}{cmyk}{0,0,1,0}
}

\usepackage{hyperref,txfonts}
\usepackage{xcolor}
\hypersetup{
    colorlinks,
    citecolor=blue,
    filecolor=blue,
    linkcolor=blue,
    urlcolor=blue,
    pdfstartview=FitH
}

\makeatother

\usepackage{babel}
\begin{document}

\title{Bulk disorder in the superconductor affects proximity-induced topological
superconductivity}

\author{Hoi-Yin Hui}

\affiliation{Department of Physics, Condensed Matter Theory Center and Joint Quantum
Institute, University of Maryland, College Park, Maryland 20742-4111,
USA}

\author{Jay D. Sau}

\affiliation{Department of Physics, Condensed Matter Theory Center and Joint Quantum
Institute, University of Maryland, College Park, Maryland 20742-4111,
USA}

\author{S. Das Sarma}

\affiliation{Department of Physics, Condensed Matter Theory Center and Joint Quantum
Institute, University of Maryland, College Park, Maryland 20742-4111,
USA}

\date{\today}
\begin{abstract}
We investigate effects of ordinary nonmagnetic disorder in the bulk
of a superconductor on magnetic adatom-induced Shiba states and on
the proximity-induced superconductivity in a nanowire that is tunnel
coupled to the bulk superconductor. Within the formalism of self-consistent
Born approximation, we show that, contrary to widespread belief, the
proximity-induced topological superconductivity can be adversely affected
by the bulk superconducting disorder even in the absence of any disorder
in the nanowire (or the superconductor-nanowire interface) when the
proximity tunnel coupling is strong. In particular, bulk disorder
can effectively randomize the Shiba-state energies. In the case of
a proximate semiconductor nanowire, we numerically compute the dependence
of the effective disorder and pairing gap induced on the wire as a
function of the semiconductor-superconductor tunnel coupling. We find
that the scaling exponent of the induced disorder with respect to
coupling is always larger than that of the induced gap, implying that
at weak coupling, the proximity-induced pairing gap dominates, whereas
at strong coupling, the induced disorder dominates. These findings
bring out the importance of improving the quality of the bulk superconductor
itself (in addition to the quality of the nanowire and the interface)
in the experimental search for solid-state Majorana fermions in proximity-coupled
hybrid structures and, in particular, points out the pitfall of pursuing
strong coupling between the semiconductor and the superconductor in
a goal toward having a large proximity gap. In particular, our work
establishes that the bulk superconductor in strongly coupled hybrid
systems for Majorana studies must be in the ultraclean limit, since
otherwise the bulk disorder is likely to completely suppress all induced
topological superconductivity effects.
\end{abstract}


\maketitle

\section{Introduction\label{sec:Introduction}}

Majorana fermions in solid-state systems \cite{Stanescu2013,Alice2012,Leijnse2012,Beenakker2013,Elliott2015}
obey non-Abelian braiding statistics \cite{Nayak2008} and are a promising
platform for topological quantum computation \cite{DasSarma2015a}.
A feasible route towards realizing them utilizes a hybrid structure
involving the proximitization of a semiconductor (SM) with a bulk
$s$-wave superconductor (SC) \cite{Lutchyn2010,Sau2010,Sau2010a,Oreg2010,Alicea2010,Lutchyn2011,Stanescu2011}.
With the appropriate combination of spin-orbit coupling (SOC), Zeeman
spin splitting, and SC pairing terms, the proximitized system becomes
topological (i.e., an effectively spinless $p$-wave superconductor)
and localized Majorana fermions emerge at the ends of one-dimensional
(nanowire) systems or at the vortices of two-dimensional systems.
Their existence can then be probed by conductance measurements as
quantized zero-bias peaks of height $2e^{2}/h$ at zero temperature
associated with the perfect Andreev reflection induced by the Majorana
zero energy modes \cite{Sengupta2001,Law2009,Sau2010a,Flensberg2010,Wimmer2011,Fregoso2013}.
Shortly after the theoretical proposals \cite{Lutchyn2010,Sau2010,Sau2010a,Oreg2010,Alicea2010,Lutchyn2011,Potter2011b}
were put forward, several experimental groups implemented different
variants of the proposed Majorana experiment using nanowires in proximity
to bulk superconductors \cite{Mourik2012,Das2012,Deng2012,Rokhinson2012,Finck2013,Churchill2013,Nadj-Perge2014}.
Although the initial data reporting zero-bias tunneling conductance
peaks in nanowires (albeit with conductance values below the theoretically
predicted $2e^{2}/h$ quantized conductance) are encouraging, more
theoretical and experimental work still needs to be done in order
to distinguish signatures of Majorana fermions from those from other
possible nontopological mechanisms as have been discussed in the literature
\cite{Motrunich2001,Liu2012,Sau2013,Hui2014a,Pikulin2012,Sau2015,Dumitrescu2015,DasSarma2015}.

The current work is on the deleterious effect of disorder on the proximity-induced
topological superconductivity in the hybrid system of experimental
interest. The topological superconductivity induced in the nanowire,
arising from a combination of $s$-wave superconductivity, spin splitting,
and spin-orbit coupling, is essentially equivalent to a type of an
effectively spinless $p$-wave superconductivity \cite{Tewari2010}
with triplet spin correlations \cite{Liu2015} which is not immune
to ordinary nonmagnetic disorder in the environment, unlike regular
\emph{s}-wave spin-singlet superconductors which are protected against
nonmagnetic disorder by virtue of the Anderson theorem. There have
therefore been many theoretical and numerical studies \cite{Motrunich2001,Akhmerov2011,Brouwer2011,Brouwer2011a,Stoudenmire2011,Bagrets2012,Liu2012,Lobos2012,Sau2012a,Asano2013,DeGottardi2013,Neven2013,Sau2013,Tkachov2013,Vayrynen2013,Adagideli2014,Beenakker2014,Hui2014,Stanev2014,Keser2015}
of the effects of disorder on the topological superconductivity in
this context, going back to almost 15 years ago \cite{Motrunich2001}.
It may appear that another theoretical study of disorder effects in
this context would be redundant, but as we explain below, this is
not the case here. The specific question regarding disorder effects
(in the bulk superconductor itself) addressed in this paper has only
been discussed a few times in the literature, with the first paper
\cite{Potter2011} coming to an erroneous conclusion which was subsequently
corrected \cite{Potter2011a,Lutchyn2012}. The conclusion we reach
in our current work is of \emph{great} importance in choosing the
proper materials for the hybrid structures manifesting topological
superconductivity and Majorana fermions.

In considering the effects of disorder in these hybrid SM-SC \cite{Mourik2012}
(or FM-SC where FM stands for the ferromagnetic adatoms as in Ref.~\cite{Nadj-Perge2014})
systems, one should distinguish between disorder in the SM and that
in the bulk SC. Disorder in the SM has been extensively studied \cite{Brouwer2011,Brouwer2011a,Sau2012a,Sau2013,DeGottardi2013,Neven2013},
with the main conclusion being that the topological gap is destroyed
by this type of disorder when the mean free path is comparable with
or smaller than the induced coherence length in the semiconductor.
It has been much emphasized in the literature \cite{Sau2012a} that
the topological nanowire must be in the ballistic limit with the carrier
mean free path being much larger than the proximity-induced coherence
length in the nanowire for the manifestation of the Majorana zero
modes, a condition which is likely (unlikely) to be satisfied in the
semiconductor \cite{Mourik2012} (ferromagnetic \cite{Nadj-Perge2014})
nanowires. It has also been emphasized \cite{Takei2013} that the
applicable disorder at the nanowire-superconductor interface must
be low for the induced proximity superconductivity to manifest a hard
gap as has recently been reportedly accomplished in the InAs-Al epitaxial
core shell nanowire hybrid structures \cite{Chang2015}. On the other
hand, disorder in the bulk SC has received relatively little attention
\cite{Potter2011,Potter2011a,Lutchyn2012,Stanescu2011}, with the
focus mainly on the limit where the coupling between the two materials
is small (i.e. the weak-coupling limit where the SM-SC tunneling amplitude
is small). In this limit, it has been found \cite{Lutchyn2012,Potter2011}
that the disorder in the bulk SC hardly affects the superconducting
gap in the topological system, making it possible to use disordered
or dirty SCs in experiments \cite{Mourik2012,Das2012,Deng2012,Rokhinson2012,Finck2013}.
One consensus in the community regarding disorder effects seems to
be that disorder in the nanowire itself (superconductor itself) is
important (unimportant) with respect to the manifestation of proximity-induced
topological superconductivity and Majorana fermions in the hybrid
system. The current work directly challenges this consensus, showing
that the disorder in the bulk superconductor may very well be important
for the proximity-induced topological superconductivity, particularly
in the limit where the superconductor and the nanowire are strongly
tunnel coupled. In particular, the bulk superconductor should be in
the clean limit, with its elastic mean free path being much larger
than the superconducting coherence length for optimal induced topological
superconducting order in the hybrid structures. \emph{Our current
work indicates that having a clean superconductor with a very long
mean free path is an absolute necessary condition for the realization
of a large proximity-induced topological superconducting gap hosting
Majorana fermions in the SM-SC and FM-SC hybrid systems.}

Two recent developments prompted us to revisit the issue of bulk disorder
in the superconductor. First, a new class of proposals \cite{Lee2009,Choy2011,Chung2011,Duckheim2011,Nadj-Perge2013,Nakosai2013,Pientka2013,Nadj-Perge2014,Pientka2014,Poyhonen2014,Brydon2015,Hui2015,Heimes2015,Weststroem2015,Peng2015},
utilizing Shiba states induced by magnetic adatoms on SCs to generate
Majorana fermions, has emerged. This platform for using ferromagnetic
adatoms on a superconducting substrate as the topological FM-SC hybrid
system is, in some sense, the large spin splitting limit \cite{Dumitrescu2015}
of the SM-SC hybrid structure with the spin-splitting arising intrinsically
from exchange effects in the ferromagnet, rather than from a Zeeman
splitting induced by an external magnetic field as in the SM-SC hybrid
system. This ferromagnet-superconductor hybrid system can therefore
be effectively described by a Hamiltonian which is the same as that
of the SM-SC heterostructure, with a crucial difference that the tunnel
coupling between the adatoms and the SC (as well as the spin splitting
in the adatom chain) is much larger than the corresponding term in
the SM-SC system \cite{DasSarma2015}, rendering the previous perturbative
treatment of disorder in SC inapplicable. Second, in Ref.~\cite{Chang2015},
the SM-SC structure has been grown epitaxially, which drastically
improved the quality of the interface between the two materials. A
hard proximity-induced superconducting gap is then observed on the
SM, resolving the ``soft-gap'' issue that previous experiments found
\cite{Takei2013}. The size of the gap on SM is comparable to that
on the SC, indicating strong coupling between the two materials \cite{Sau2010c}.
In this limit, however, it is unclear whether disorder in the bulk
SC can significantly degrade the gap on the SM, especially when a
magnetic field is applied on the SM to create a Zeeman spin splitting
necessary for producing the topological superconductivity in the SM
wire.

In both of these experiments \cite{Nadj-Perge2014,Chang2015}, the
strong tunnel coupling between the SM (or magnetic adatoms) and the
SC necessitates the re-examination of the issue of disorder in the
SC, as previous treatments of this problem were valid only when the
coupling is small \cite{Potter2011,Lutchyn2012,Potter2011a}, which
is not the case in these two new systems. In this paper, we investigate
the effects of disorder in a SC on the spectral properties of a proximate
SM in the SM-SC system and on the Shiba states in the FM-SC system.
The disorder problems for the two hybrid structures (i.e., disorder
effects on the Shiba states in the ferromagnetic adatom chain and
on the SC/SM nanowire) are somewhat different, and we therefore study
the two systems (FM/SC and SM/SC) separately so that our work applies
to both experimental systems, although our main emphasis in the current
work is on the semiconductor-based Majorana hybrid systems since the
experimental situation is better understood in such semiconductor-nanowire
structures. The formalism we adopt is the self-consistent Born approximation,
which is valid in the limit of weak impurity scattering (specifically
$k_{F}l\gg1$, where $k_{F}$ and $l$ are, respectively, the Fermi
wave number and the disorder-induced transport mean free path in the
bulk SC- this condition is well satisfied in the bulk superconductors
used in the Majorana hybrid structures with the clean/dirty bulk superconductors
being defined by whether $l\gg\xi$ or $l\ll\xi$, respectively, where
$\xi$ is the SC coherence length). We extract the density of states
(DOS) of the topological systems via their dynamic Green functions
which contain the self-energy due to ensemble-averaged disorder in
the bulk of the SC. Throughout the paper, we shall assume the SM itself
as well as the SM-SC interface is clean and only consider disorder
in the bulk superconductor.

Investigation of whether a strong tunnel coupling to the superconductor
in hybrid systems, e.g. the inherent strong coupling in metal-on-metal
ferromagnetic adatom systems or the strong coupling in epitaxial semiconductor-superconductor
systems, could lead to the disorder in the bulk superconductor (quite
apart from the disorder in the adatoms or the semiconductor itself)
becoming relevant for the topological superconducting (and consequently
Majorana fermion) properties is the goal of the current theoretical
work. We ignore all disorder in the SM nanowire (or the FM chain)
itself since it has already been studied extensively elsewhere and
is well understood as being detrimental to topological properties.

To qualitatively understand how the bulk superconducting disorder
might be relevant for the proximity-induced superconductivity in the
strong tunnel coupling (which we often refer to simply as ``strong
coupling'' in this paper) limit, we refer to the simple (and approximate)
formula for the proximity-induced superconducting gap in the hybrid
system derived in Refs.~\cite{Sau2010a,Sau2010c,Sau2012a} and used
extensively:
\begin{equation}
\Delta_{w}\sim\frac{\Gamma\Delta}{\Gamma+\Delta}
\end{equation}
where $\Gamma$ is the effective coupling, $\Delta$ is the bulk gap
in the superconductor, and $\Delta_{w}$ is the induced proximity
gap in the nanowire. It has been much emphasized that in order to
obtain a large induced gap, one must have $\Gamma\gg\Delta$ so that
$\Delta_{w}\sim\Delta$, which is obviously the maximum possible value
of the induced gap (as achieved presumably in the epitaxial InAs-Al
core-shell nanowire systems \cite{Chang2015}). In the opposite limit
of very weak coupling, $\Gamma\ll\Delta$, one gets $\Delta_{w}\sim\Gamma$
with a very small induced gap $\ll\Delta$ (with consequently an even
smaller topological gap since the topological gap is bounded from
above by $\Delta_{w}$). Let us now imagine an extremely large tunnel
coupling (e.g., $\Gamma$ going to infinity) where there is then no
discernible difference between the superconductor and the nanowire
so that $\Delta_{w}=\Delta$ applies, and hence the nanowire has essentially
become a part of the bulk superconductor as far as superconducting
properties go. In such a situation, the bulk disorder in the superconductor
is now a part of the disorder in the nanowire since from the perspective
of superconductivity, these two have become one monolithic system.
Now, if we turn on spin-orbit coupling and spin splitting so as to
convert $\Delta_{w}$ into a topological superconducting gap, then
the disorder existing in the bulk superconductor must necessarily
suppress the effectively triplet spinless topological superconductivity
since it is not protected by any Anderson theorem (as time reversal
invariance is explicitly broken). We note that this argument does
not apply in the weak tunneling ($\Gamma\ll\Delta$) limit where the
two parts of the hybrid system (the bulk superconductor and the nanowire)
are distinct, and indeed it has been explicitly shown \cite{Lutchyn2012}
that in the weak tunneling limit ($\Gamma$ going to zero), the bulk
superconducting disorder does not suppress the topological superconductivity,
but of course the topological gap is very small in this limit ($<\Gamma$)
anyway. Although this physically motivated qualitative argument is
not a proof by any means, the argument demonstrates that the strong-
and weak-coupling situations could be fundamentally different with
respect to disorder effects coming from the bulk superconductor, and
a careful investigation is necessary to see whether or not the strong-coupling
situation is benign with respect to the bulk disorder effects. While
this problem is of considerable intrinsic interest itself in the context
of the theory of superconducting proximity effect, the current experimental
push (e.g., InAs-Al epitaxial hybrid system) to produce a hard induced
gap makes our work timely in the study of Majorana fermions in solid-state
systems.

The paper is organized as follows. In Sec.~\ref{sec:Shiba}, we consider
the ferromagnet-SC hybrid system in the Shiba limit, where the coupling
between the magnetic adatoms and the SC is much stronger than the
interatomic coupling. We find that in this strong-coupling regime,
disorder in the bulk SC has strong effects on the location of the
Shiba energy in the bulk SC gap. In Sec.~\ref{sec:Nanowire}, we
consider the effects of bulk superconducting disorder on the SM-SC
heterostructure, in both the weak- and strong-coupling limits. We
show that our results in the weak-coupling limit agree with previous
works \cite{Potter2011,Lutchyn2012}, and highlight features specific
to the strong-coupling limit, where, in contrast to the weak-coupling
limit, nonmagnetic elastic disorder in the bulk superconductor invariably
strongly degrades the proximity-induced topological SM superconductivity.
We conclude in Sec.~\ref{sec:Conclusion} with a summary and with
a brief discussion of the far-reaching implications of our findings
for the future design of hybrid structures hosting Majorana fermions.

\section{Ferromagnetic Adatom-Induced Shiba States in a Disordered Superconductor\label{sec:Shiba}}

We first consider a disordered $s$-wave SC strongly coupled with
a magnetic impurity, described by
\begin{eqnarray}
H & = & \sum_{\bm{k}\sigma}\xi_{\bm{k}}a_{\bm{k}\sigma}^{\dagger}a_{\bm{k}\sigma}+\Delta\sum_{\bm{k}}\left(a_{\bm{k}\uparrow}^{\dagger}a_{\bm{k}\downarrow}^{\dagger}+{\rm h.c.}\right)\nonumber \\
 &  & -J\sum_{\sigma}\sigma a_{\sigma}^{\dagger}\left(\bm{r}=0\right)a_{\sigma}\left(\bm{r}=0\right)\label{eq:H}\\
 &  & +\int d\bm{r}U_{{\rm dis}}\left(\bm{r}\right)\sum_{\sigma}a_{\sigma}^{\dagger}\left(\bm{r}\right)a_{\sigma}\left(\bm{r}\right),\nonumber 
\end{eqnarray}
where $a_{\bm{k}\sigma}$ annihilates an electron with momentum $\bm{k}$
and spin $\sigma$. In the first line, $\xi_{\bm{k}}$ is the normal-state
dispersion and $\Delta$ is the $s$-wave pairing term of the SC.
In the second line, $J$ characterizes the strength of the magnetic
impurity, located at the origin $\left(\bm{r}=0\right)$, which induces
a local Zeeman term in the SC. The prefactor $\sigma=\pm1$ corresponds
to spin-up/down respectively. In the third line, $U_{{\rm dis}}\left(\bm{r}\right)$
represents nonmagnetic elastic disorder present in the SC (which leads
to a finite transport mean free path $l$ in the bulk SC in its normal
state). Below, we investigate effects of the magnetic term $\left(J\right)$
on the DOS of the system, for both clean $\left(U_{{\rm dis}}=0\right)$
and dirty $\left(U_{{\rm dis}}\neq0\right)$ SCs by generalizing the
original Yu-Shiba-Rusinov theory \cite{Yu1965,Shiba1968,Rusinov1968}
to include static nonmagnetic elastic disorder $U_{{\rm dis}}$ in
the SC.

\subsection{Clean Superconductor}

\begin{figure}
\begin{centering}
\includegraphics[width=0.75\columnwidth]{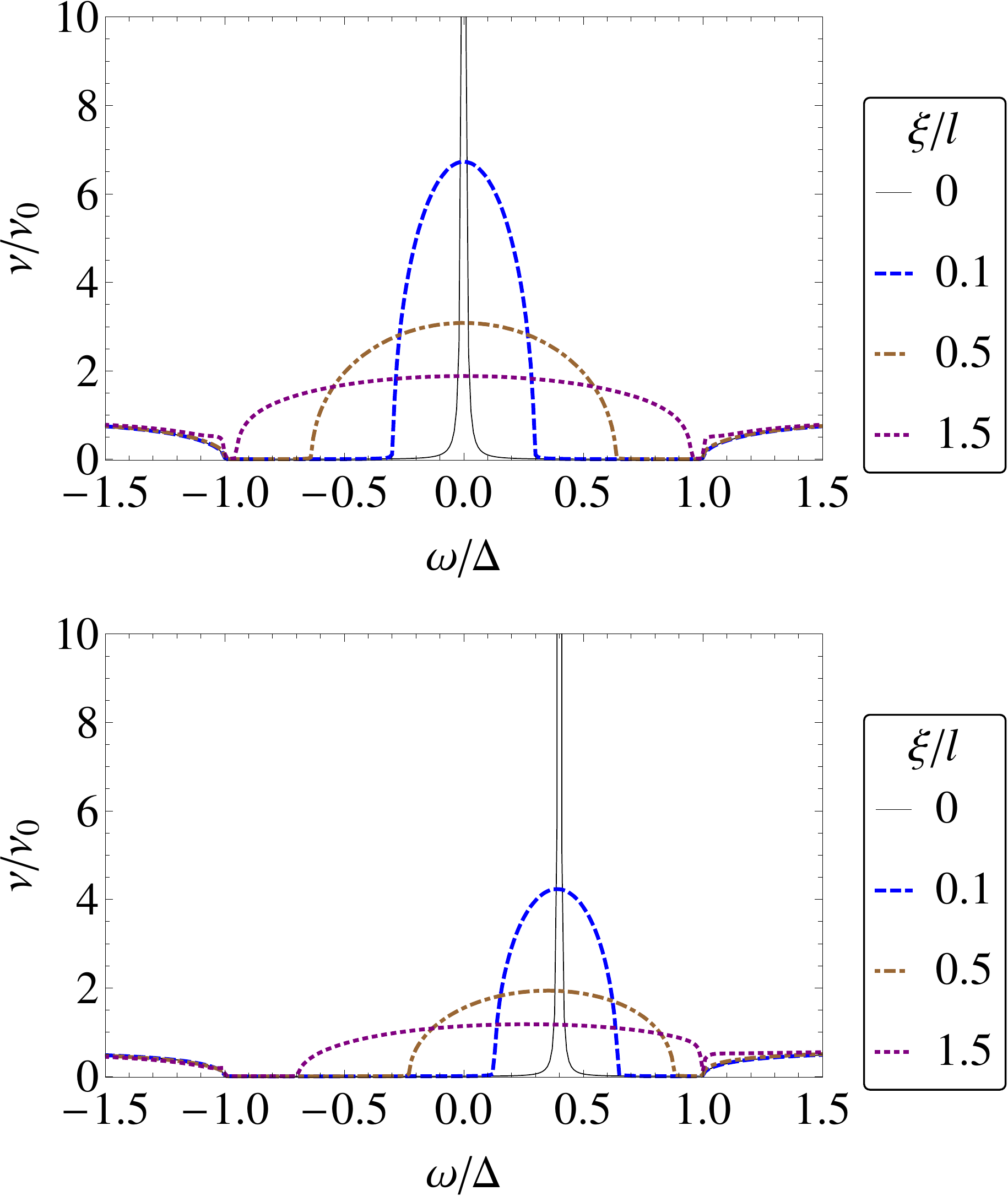}
\par\end{centering}

\caption{(Color online) The LDOS at the position of the magnetic impurity in
a clean (black solid lines) or disordered (colored lines) SC. The
Shiba-state energies are tuned to (a) $\epsilon_{0}=0$ and (b) $\epsilon_{0}=0.4\Delta$.
A broadening of magnitude $0.001\Delta$ is used to smear out the
$\delta$-functions for depicting the results. The elastic disorder
in the system is quantified by the mean free path $l$, which is given
in the units of the clean limit coherence length $\xi$ of the SC,
as shown on the right. \label{fig:LDOS}}
\end{figure}

We first briefly review the theory of Shiba states \cite{Yu1965,Shiba1968,Rusinov1968}
in the absence of disorder $\left(U_{{\rm dis}}=0\right)$ to set
a context and to fix the terminology. In frequency $\left(\omega\right)$--momentum
$\left(\bm{k}\right)$ space, the Green function for the system is
\begin{equation}
G_{\bm{kk'}}^{(1)}=G_{\bm{k}}^{(0)}\delta_{\bm{kk'}}+G_{\bm{k}}^{(0)}T_{\bm{kk'}}G_{\bm{k'}}^{(0)},\label{eq:GTG}
\end{equation}
where $G_{\bm{k}}^{(0)}\left(\omega\right)=\frac{\omega\tau_{0}+\xi_{\bm{k}}\tau_{z}+\Delta\tau_{x}}{\omega^{2}-\xi_{\bm{k}}^{2}-\Delta^{2}}$
is the Green function for a clean SC with $\tau_{\mu}$ the Pauli
matrices acting on the Nambu-Gorkov space of $\left(a_{\bm{k}\uparrow},a_{\bm{k}\downarrow}^{\dagger}\right)^{T}$.
The superscript $\left(1\right)$ in Eq.~(\ref{eq:GTG}) indicates
the presence of one magnetic impurity, but without disorder in the
system. The effect of the magnetic term $J$ is captured by the $T$
matrix in the second term of Eq.~(\ref{eq:GTG}), which is given
by
\begin{eqnarray}
T_{\bm{kk'}} & = & -\left(1+\frac{J}{V}\sum_{\bm{k}}G_{\bm{k}}^{(0)}\right)^{-1}\frac{J}{V}\nonumber \\
 & = & -\left(1-J\pi\nu_{0}\frac{\omega\tau_{0}+\Delta\tau_{z}}{\sqrt{\omega^{2}-\Delta^{2}}}\right)^{-1}\frac{J}{V},
\end{eqnarray}
where $V$ is the volume of the system and $\nu_{0}$ is the normal-state
DOS at the Fermi level. The pole of $T$ in the subgap regime indicates
the presence of a bound state, called the Shiba state \cite{Yu1965,Shiba1968,Rusinov1968},
with the energy given by
\begin{equation}
\epsilon_{0}={\rm sgn}J\frac{1-\left(J\pi\nu_{0}\right)^{2}}{1+\left(J\pi\nu_{0}\right)^{2}}\Delta.\label{eq:shibaE}
\end{equation}
The local density of states (LDOS) at the position of the Shiba state
is given by $\nu\left(\bm{r}=0\right)=V^{-1}\sum_{\bm{kk'}}G_{\bm{kk'}}^{(1)}$.
In Fig.~\ref{fig:LDOS}, the black lines show the the LDOS at a Shiba
state with energies $\epsilon_{0}=0$ and $\epsilon_{0}=0.4\Delta$.
The divergence of the LDOS at $\omega=\epsilon_{0}$ indicates that
the Shiba states have well-defined energies. We mention the obvious
fact that the Shiba-state energy $\epsilon_{0}$ is tuned by appropriately
tuning the magnetic coupling $J$, and in a given experimental setup,
$J$ would typically be fixed producing a Shiba energy according to
Eq.~(\ref{eq:shibaE}) above. Results corresponding to two situations
with $\epsilon_{0}=0$ and $0.4\Delta$ are shown in Figs.~\ref{fig:LDOS}
and \ref{fig:dJ}.

\subsection{Disordered Superconductor}

We now investigate the effects of ensemble-averaged disorder in the
bulk SC $\left(U_{{\rm dis}}\right)$ on the Shiba-state energy. To
this end, we assume that the random quenched nonmagnetic impurities
in the SC have a concentration of $n_{{\rm imp}}$, and each impurity
has a scattering potential of the form $U_{i}\left(\bm{r}\right)=U\delta\left(\bm{r}-\bm{r}_{i}\right)$,
where $\bm{r}_{i}$ is the position of the $i^{th}$ impurity. In
the self-consistent Born approximation, the Green function is written
as
\begin{equation}
G_{\bm{kk'}}=\left[\left(G_{\bm{kk'}}^{(1)}\right)^{-1}-\Sigma_{\bm{kk'}}\right]^{-1},\label{eq:Gshiba}
\end{equation}
where $G^{(1)}$ is given by Eq.~(\ref{eq:GTG}) and the inversion
here is operated on the $\bm{k}-\bm{k'}$ matrix space. The disorder-induced
self-energy $\Sigma_{\bm{kk'}}$ is given (in the self-consistent
Born approximation) by 
\begin{eqnarray}
\Sigma_{\bm{kk'}} & = & n_{{\rm imp}}U^{2}\frac{1}{V}\sum_{\bm{p}}\tau_{z}G_{\bm{p}+\bm{k},\bm{p}+\bm{k'}}\tau_{z},\label{eq:SigmaExact}\\
 & \approx & \frac{\tau^{-1}}{2\pi\nu_{0}}\delta_{\bm{k}\bm{k'}}\frac{1}{V}\sum_{\bm{pq}}\tau_{z}G_{\bm{p},\bm{q}}\tau_{z}.\label{eq:SigmaApprox}
\end{eqnarray}
where $\tau=\left(2\pi n_{{\rm imp}}U^{2}\nu_{0}\right)^{-1}$ is
the disorder scattering time. Note that due to the lack of translational
invariance, the self-energy due to disorder, in its exact form of
Eq.~(\ref{eq:SigmaExact}), is nondiagonal in $\bm{k}$--$\bm{k'}$.
It is easy to check that Eq.~(\ref{eq:SigmaExact}) reduces to the
conventional form for translationally invariant systems \cite{Mahan2000,*Abrikosov1975}
if $G_{\bm{kk'}}$ is proportional to $\delta_{\bm{kk'}}$.

In reaching Eq.~(\ref{eq:SigmaApprox}), we observed that that $G_{\bm{kk'}}^{(1)}$
has the highest weight when $\left|\bm{k}\right|=\left|\bm{k'}\right|=k_{F}^{(N)}$,
where $k_{F}^{(N)}$ is the Fermi momentum of the system in its normal
state. Therefore, in the summation over $\bm{p}$ in Eq.~(\ref{eq:SigmaExact}),
the summand has appreciable weights only when $\left|\bm{p}+\bm{k}\right|=\left|\bm{p}+\bm{k'}\right|=k_{F}^{(N)}$.
In the general case where $\bm{k}\neq\bm{k'}$, this condition is
satisfied only for a one-dimensional manifold of $\bm{p}$, but when
$\bm{k}=\bm{k'}$, the condition reduces to $\left|\bm{p}+\bm{k}\right|=k_{F}^{(N)}$
and is satisfied by a two-dimensional manifold of $\bm{p}$. Thus
we see that $\Sigma_{\bm{kk'}}$ has most of its weights at $\bm{k}=\bm{k'}$,
allowing us to approximate it by Eq.~(\ref{eq:SigmaApprox}). This
can be understood physically as the ensemble-averaged disorder should
not introduce further translational-symmetry breaking and hence $\Sigma_{\bm{kk'}}$
is diagonal in $\bm{k}$.

For each value of energy $\left(\omega\right)$, Eqs.~(\ref{eq:Gshiba})
and (\ref{eq:SigmaApprox}) are iterated numerically until convergence.
In the evaluation of the momentum integrals, we use the approximation
$\frac{1}{V}\sum_{\bm{k}}\rightarrow\nu_{0}\int_{-\infty}^{\infty}d\xi_{\bm{k}}$.
The $\xi_{\bm{k}}$ integral is discretized on a grid with $10^{4}$
points distributed in a way such that $\frac{1}{1+\left|\xi_{\bm{k}}\right|}$
is sampled uniformly over the interval $\left[10^{-4},1\right]$.
The iteration converges after a few cycles for most values of $\omega$
except for those in the vicinity of the Shiba-state energy $\left(\epsilon_{0}\right)$,
which requires a few hundred of iteration cycles.

The blue dashed line in Fig.~\ref{fig:LDOS} shows the LDOS at the
magnetic impurity when the SC is disordered with a mean free path
of $l=v_{F}\tau=10\xi$, where $v_{F}$ and $\xi=v_{F}/\Delta$ are
the Fermi velocity and the coherence length of the SC, respectively.
The results for several other values of $l$ are also presented. We
observe that the $\delta$ peak at $\omega/\Delta=\epsilon_{0}$ associated
with the Shiba state is now broadened to a dome by the disorder in
the SC. This can be understood as follows: the continuum states of
the SC are different for each realization of disorder. The scattering
phase shift due to the magnetic impurity therefore varies from one
realization to another, leading to an effective disorder-induced fluctuation
in $J$ and $\epsilon_{0}$ {[}c.f. Eq.~(\ref{eq:shibaE}){]}. Ensemble
averaging the LDOS over disorder leads to a domelike shape centered
around the energy of the Shiba state in a clean SC with the dome in
Fig.~\ref{fig:LDOS} reflecting the ``spreading'' in the effective
Shiba energy due to disorder---it is clear that the clean SC limit
with $l\gg\xi$ is necessary for the system to have a sharp Shiba
energy.

\begin{figure}
\begin{centering}
\includegraphics[width=0.6\columnwidth]{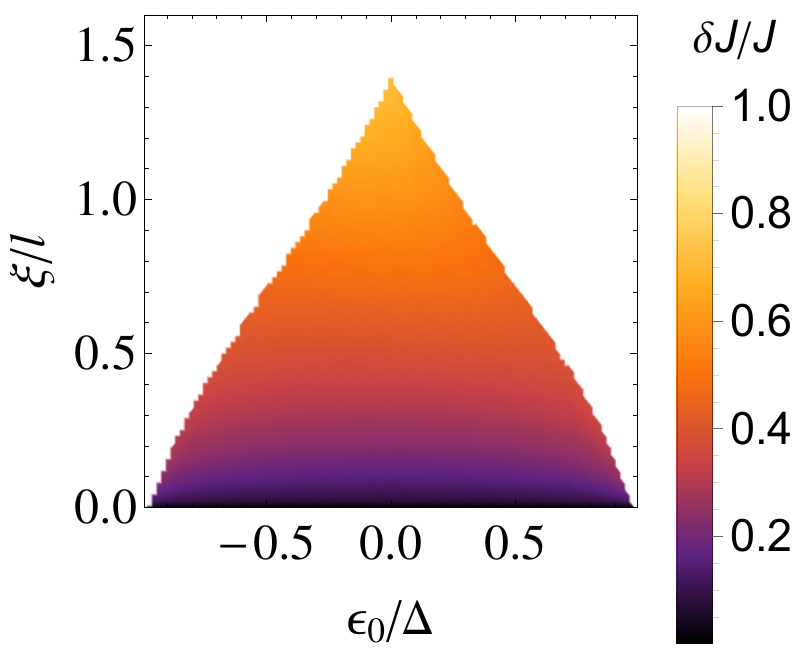}
\par\end{centering}

\begin{centering}
\includegraphics[width=0.8\columnwidth]{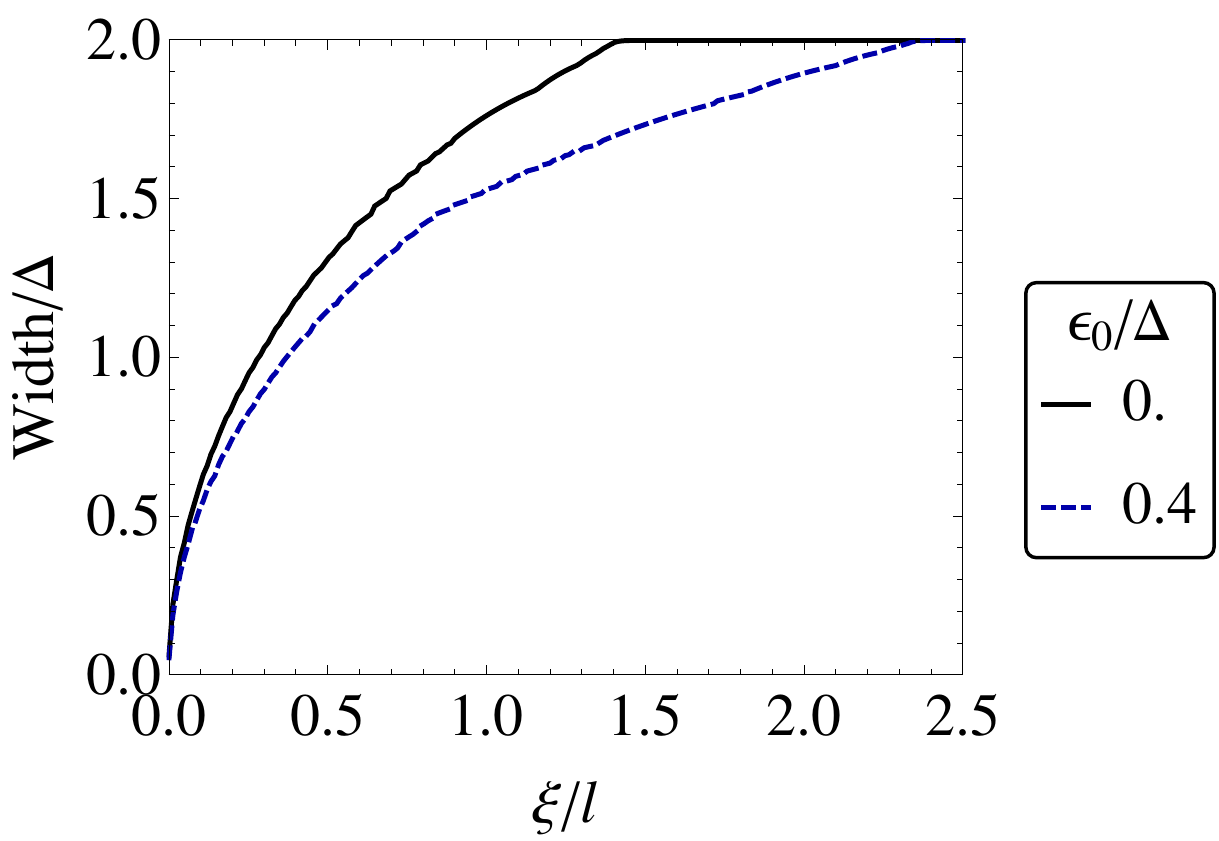}
\par\end{centering}

\caption{(Color online) (a) The normalized fluctuation $\delta J/J$ as a function
of the disorder strength $\xi/l$ and the Shiba energy $\epsilon_{0}$
for clean SC. In the white regions, the subgap dome of LDOS joins
the continuum modes $\left(\left|\omega\right|>\Delta\right)$ which
makes $\delta J$ ill defined. (b) The width of the subgap dome as
a function of disorder strength for $\epsilon_{0}=0$ (solid line)
and $\epsilon_{0}=0.4\Delta$ (dashed line).\label{fig:dJ}}
\end{figure}

We plot in Fig.~\ref{fig:dJ}(a) the effective fluctuation in $J$,
which is defined as the standard deviation $\delta J$ of the distribution
in $J$ that would result in Shiba-state energies distributed according
to the subgap LDOS obtained from the self-consistent Born approximation
(e.g., the dome in Fig.~\ref{fig:LDOS}). Fixing the Shiba-state
energy in the clean limit at $\epsilon_{0}=0$ and $0.4\Delta$ (as
in Fig.~\ref{fig:LDOS}), we plot in Fig.~\ref{fig:dJ}(b) the width
of the subgap dome of LDOS against the strength of disorder as characterized
by $\xi/l$. In general, the Shiba subgap state is broadened by disorder,
and the subgap DOS joins the continuum states when $l\lesssim\xi$,
with the precise critical disorder dependent on the Shiba energy.
The results presented in Figs.~\ref{fig:LDOS} and \ref{fig:dJ}
clearly demonstrate the importance of being in the ultraclean-SC limit,
i.e., $l\gg\xi$, for obtaining a Shiba state whose energy is close
to that in the clean limit.

\subsection{Discussion}

We have investigated the LDOS associated with a single magnetic impurity
embedded in a disordered SC with ensemble averaging. The result thus
obtained is \emph{not} expected to be directly applicable to a real
experiment conducted with a single impurity since in reality there
is only one realization of disorder, and therefore this Shiba-state
energy should appear as a sharp subgap LDOS peak. Our ensemble-averaged
results, however, reveal that the Shiba-state energy is shifted randomly
from sample to sample around its value in the clean-SC limit (with
the likelihood roughly proportional to the height of the subgap dome)
because each specific experimental sample will have its unique disorder
configuration which will differ randomly from one sample to another.
Qualitatively, when the mean free path of the bulk SC is of the same
order of magnitude as (or shorter than) its SC coherence length, the
Shiba-state energy could be anywhere within the SC gap. This is likely
to have implications for a class of recent proposals which utilize
the Shiba states induced in a superconductor by a chain of magnetic
adatoms to generate Majorana fermions \cite{Choy2011,Pientka2013,Nadj-Perge2013,Pientka2014,Nadj-Perge2014,Brydon2015,Weststroem2015,Peng2015}.
In general, the spin-orbit-coupling strength, the lattice spacing
between magnetic adatoms, and the Shiba-state energy all need to be
fine tuned to obtain topological superconductivity in the system \cite{Pientka2014,Weststroem2015}.
However, it is difficult to control the Shiba-state energy even if
the SC is clean, as it is determined by the strengths of the magnetic
adatom and its coupling with the SC, both of which can hardly be tuned
experimentally. Our finding indicates that in addition to the difficult
task of fine tuning the Shiba energy to zero, it is also necessary
to use an ultraclean SC with $l\gg\xi$ so as to ensure that the Shiba-state
energy for each individual atom remain close to zero. Otherwise, with
Shiba-state energies along the chain being shifted randomly by disorder,
the system would then be fragmented into segments of topological and
nontopological regions, an unfavorable situation for topological quantum
computation. Thus, our current work implies that the fine-tuning problem
of creating Majorana modes using Shiba states becomes substantially
worse in the strong-coupling situation since the bulk disorder in
the superconductor now randomly shifts the Shiba-state energy, leading
to strong sample-to-sample variations. We mention here that the typical
Shiba-induced Majorana system is a metal-on-metal system (i.e., a
ferromagnetic metal chain on a superconducting metal) where the tunnel
coupling is large ($\sim$eV), and the typical bulk disorder scale
in the superconductor ($\sim$meV) is much larger than the typical
induced gap ($\sim0.1$ meV), leading to an intrinsically unfavorable
theoretical situation for the existence of Majorana modes by virtue
of the fine-tuning problem.

\section{Semiconductor Nanowire-Superconductor System\label{sec:Nanowire}}

In Sec.~\ref{sec:Shiba}, we looked into the system in the ``Shiba
limit,'' in which the interatomic hopping among the magnetic adatoms
is much weaker than their coupling with the bulk SC. We now turn to
consider the opposite limit where the system hybridizes strongly to
form a nanowire. (We mention as an aside that the two limiting situations,
i.e., the Shiba limit of weak interwire hopping \cite{Brydon2015}
and the nanowire limit of strong interwire hopping \cite{Hui2015},
are not separated by a quantum phase transition and are two extremes,
which are applicable to different physical situations, of the same
underlying physics \cite{Peng2015,DasSarma2015}.) Previous work \cite{Lutchyn2012}
has indicated that disorder  in the bulk SC cannot degrade the superconducting
gap in the nanowire \emph{if }the SM-SC coupling is weak. It, however,
remains unclear whether a strong SM-SC coupling could alter this conclusion
qualitatively (as was already discussed in Sec.~\ref{sec:Introduction}
of this paper). Therefore, we now investigate this question in depth
without assuming any weak SM-SC coupling. Note that the effect of
disorder at the SM-SC interface is a separate issue which has been
theoretically treated previously \cite{Takei2013}.

In the absence of disorder, the system is described by the Hamiltonian
$H=H_{w}+H_{sc}+H_{T}$, where $H_{w/sc}$ is the Hamiltonian for
the SM wire/SC and $H_{T}$ is the coupling between the two materials.
Explicitly, they are given by \begin{subequations}
\begin{align}
H_{w} & =\sum_{k_{z}\sigma}\left[\xi_{k_{z}}^{(w)}c_{k_{z}\sigma}^{\dagger}c_{k_{z}\sigma}+B\sigma c_{k_{z}\sigma}^{\dagger}c_{k_{z}\sigma}+\alpha k_{z}c_{k_{z}\bar{\sigma}}^{\dagger}c_{k_{z}\sigma}\right]\\
H_{sc} & =\sum_{\bm{k}\sigma}\xi_{\bm{k}}^{(s)}a_{\bm{k}\sigma}^{\dagger}a_{\bm{k}\sigma}+\Delta\sum_{\bm{k}}\left(a_{\bm{k}\uparrow}^{\dagger}a_{\bm{k}\downarrow}^{\dagger}+{\rm h.c.}\right)\\
H_{T} & =\sum_{\bm{k}\sigma}a_{\bm{k}\sigma}^{\dagger}c_{k_{z}\sigma}+{\rm h.c.}
\end{align}
\end{subequations}where $\xi_{k_{z}}^{(w)}=k_{z}^{2}-\mu$ and $\xi_{\bm{k}}^{(s)}$
are the dispersions of the wire and the SC, respectively. $B$ and
$\alpha$ are the Zeeman and SOC terms on the wire, and $\Delta$
is the $s$-wave pairing term on the SC. The subscript $\bar{\sigma}$
represents the spin species opposite to that of $\sigma$. The physical
properties of the system are captured by its Green function, which
is given by
\begin{equation}
{\cal G}=\left(\begin{array}{cc}
G_{s}^{(0)-1}-\Sigma_{{\rm dis}} & T\\
T^{\dagger} & G_{w}^{(0)-1}
\end{array}\right)^{-1}\equiv\left(\begin{array}{cc}
G_{s}\\
 & G_{w}
\end{array}\right).\label{eq:Total_G}
\end{equation}
Here, the full Green function ${\cal G}$ of the whole system is written
in the block spinor space of $\left(\psi^{(s)},\psi^{(w)}\right)$,
where $\psi^{(s)/(w)}$ represents the Bogoliubov-de Gennes spinors
for the SC/wire, respectively. The off-diagonal block matrix $T$
represents the tunneling between the two systems. $\Sigma_{{\rm dis}}$
is the self-energy originating from the nonmagnetic disorder, which
is present in the SC only in our model. The explicit forms for these
terms are\begin{subequations}
\begin{align}
G_{s}^{(0)}\left(\omega,\bm{k}^{(s)}\right) & =\frac{\omega\tau_{0}+\xi_{\bm{k}}^{(s)}\tau_{z}+\Delta\tau_{x}}{\omega^{2}-\xi_{\bm{k}}^{(s)2}-\Delta^{2}},\\
G_{w}^{(0)}\left(\omega,k_{z}^{(w)}\right) & =\left(\omega\tau_{0}-\xi_{k_{z}}^{(w)}\tau_{z}-B\sigma_{z}-\alpha k_{z}^{(w)}\sigma_{x}\tau_{z}\right)^{-1},\\
T\left(\bm{k}^{(s)},k_{z}^{(w)}\right) & =t\delta_{k_{z}^{(s)},k_{z}^{(w)}},\\
\Sigma_{{\rm dis}}\left(\bm{k}^{(s)},\bm{p}^{(s)}\right) & \approx\delta_{\bm{k}^{(s)},\bm{p}^{(s)}}\frac{\tau^{-1}}{2\pi\nu_{0}V}\sum_{\bm{q}_{1},\bm{q}_{2}}G_{s}\left(\bm{q}_{1}^{(s)},\bm{q}_{2}^{(s)}\right),
\end{align}
\end{subequations}where $G_{w/s}^{(0)}$ are the unperturbed Green
functions of the wire/SC. The superscripts $\left(w\right)/\left(s\right)$
on the momentum variables indicate that they refer to the wire/SC.
The effects of disorder in the bulk SC are captured by the self-energy
$\Sigma_{{\rm dis}}$, whose expression has been approximated in the
same way as Eq.~(\ref{eq:SigmaApprox}).

Inverting the matrix in Eq.~(\ref{eq:Total_G}), we get the following
set of coupled equations:\begin{subequations}\label{eqs:GwireSC}
\begin{eqnarray}
G_{w} & = & \left(G_{w}^{(0)-1}-\Sigma_{w}\right)^{-1},\label{eq:Gwire}\\
\Sigma_{w} & = & \frac{t^{2}}{A}\sum_{\bm{k}_{\perp}}\tau_{z}\tilde{G}_{S}\left(\bm{k}\right)\tau_{z},\label{eq:withA}\\
\tilde{G}_{s} & = & \left(G_{s}^{(0)-1}-\Sigma_{{\rm dis}}\right)^{-1},\\
\Sigma_{{\rm dis}}\left(\bm{k},\bm{k'}\right) & \approx & \delta_{\bm{kk'}}\frac{\tau^{-1}}{2\pi\nu_{0}V}\sum_{\bm{q}_{1},\bm{q}_{2}}\tau_{z}G_{s}\left(\bm{q}_{1},\bm{q}_{2}\right)\tau_{z},\\
G_{s}\left(\bm{k},\bm{k'}\right) & = & \tilde{G}_{S}\delta_{\bm{kk'}}+\tilde{G}_{s}\left(\bm{k}\right)t^{2}\tau_{z}\times\nonumber \\
 &  & G_{w}\left(k_{z}\right)\tau_{z}\tilde{G}_{s}\left(\bm{k'}\right)\delta_{\bm{k}_{z},\bm{k}'_{z}},\label{eq:GS}
\end{eqnarray}
\end{subequations}where $\tilde{G}_{s}$ is the Green function of
the SC with the effects of disorder incorporated, while $G_{s}$ incorporates
both the effect of disorder and that of the coupling to the wire.
{[}$A$ in Eq.~(\ref{eq:withA}) is the area required for normalization
similar to the volume $V$ normalization in the earlier equations.{]}

The DOS on the wire can be calculated from $G_{w}$ by $\nu_{w}\left(\omega\right)=\frac{-1}{\pi}{\rm TrIm}\int\frac{dk_{z}}{2\pi}G_{w}\left(k_{z}\right)$.
To evaluate the momentum integrals, we use the following approximations\begin{subequations}
\begin{align}
\frac{1}{A}\sum_{\bm{k}_{\perp}} & \rightarrow\nu_{{\rm 2D}}\int d\xi_{\bm{k}_{\perp}}\\
\frac{1}{V}\sum_{\bm{k}} & \rightarrow\nu_{{\rm 2D}}\int_{-\infty}^{\infty}\frac{dk_{z}}{2\pi}\int d\xi_{\bm{k}_{\perp}}
\end{align}
\end{subequations}where $\nu_{2D}=\nu_{0}\pi/k_{F}$ is the density
of states of a two-dimensional system and $k_{F}$ is the Fermi momentum.

We are primarily interested in the spectral properties of the SM wire,
which becomes topological and hosts Majorana fermions with suitable
combinations of SOC, Zeeman and induced SC pairing terms. The influence
of the SC on the SM wire is captured by the self-energy term $\Sigma_{w}$.
To have a better understanding, we expand it at small frequencies
as
\begin{equation}
\Sigma_{w}\left(\omega\right)\approx\left(\Sigma_{0}+\omega\Sigma_{0}'\right)+\Sigma_{x}\tau_{x}+\ldots,\label{eq:SigmaExpansion}
\end{equation}
where $\Sigma_{\mu}=\frac{1}{4}{\rm Tr}\left[\Sigma_{w}\left(0\right)\tau_{\mu}\right]$
and $\Sigma_{0}'=\frac{1}{4}{\rm TrRe}\Sigma_{w}'\left(0\right)$
are scalar numbers, and the symbol ``$\ldots$'' represents terms
proportional to other matrices. These terms renormalize the SOC, Zeeman
splitting, effective mass, and chemical potential of the SM wire,
and do not concern us here. By substituting Eq.~(\ref{eq:SigmaExpansion})
into Eq.~(\ref{eq:Gwire}), we obtain the following form of Green
function for the wire:
\begin{align}
G_{w} & =\left[\omega-\mathcal{H}_{w}-\left(\Sigma_{0}+\omega\Sigma_{0}'\right)-\Sigma_{x}\tau_{x}\right]^{-1},\\
 & =Z^{-1}\left(\omega+\frac{i}{\tau_{w}}-Z^{-1}\mathcal{H}_{w}-Z^{-1}\Sigma_{x}\tau_{x}\right)^{-1},
\end{align}
where $Z=1-\Sigma_{0}'$ is the frequency renormalization factor,
and
\begin{equation}
\tau_{w}^{-1}=-Z^{-1}{\rm Im}\Sigma_{0}\label{eq:disorderDef}
\end{equation}
is the broadening induced by $\Sigma_{w}$. Since $\tau_{w}^{-1}$
has the same effect as disorder, we \emph{define} it as the effective
disorder on the SM wire which can be thought of as the proximity-induced
effective disorder arising in the SM due to the presence of the SC.
The term $Z^{-1}\Sigma_{x}$ induces a SC pairing onto the wire, but
we shall not define it as the pairing term directly since Eq.~(\ref{eq:SigmaExpansion})
is an expansion at zero frequency and does not capture the frequency
dependence of $\Sigma_{x}$. Rather, we numerically compute the spectral
gap when the wire has no Zeeman splitting $\left(B=0\right)$, and
identify this gap as the effective magnitude of the pairing term $\left(\Delta_{w}\right)$
in the SM wire.

Before presenting the full numerical results, we review the conventional
treatment of the problem in the weak-coupling limit where $\Gamma\ll\Delta$,
with $\Gamma=\pi\nu_{2D}t^{2}$ being the tunnel coupling strength
between the wire and the SC. In this limit, the second term of Eq.~(\ref{eq:GS})
can be neglected, and Eq.~(\ref{eqs:GwireSC}) has the analytic solution
\cite{Abrikosov1961,Maki1969}
\begin{eqnarray}
G_{s} & = & \frac{\tilde{\omega}\tau_{0}+\xi_{\bm{k}}\tau_{z}+\tilde{\Delta}\tau_{x}}{\tilde{\omega}^{2}-\xi_{\bm{k}}^{2}-\tilde{\Delta}^{2}}\\
\Sigma_{w}\left(k_{z}\right) & = & -\Gamma\frac{\omega-\Delta\tau_{x}}{\sqrt{\Delta^{2}-\omega^{2}}}\label{eq:cleanInducedGap}
\end{eqnarray}
where $\tilde{\omega}$ and $\tilde{\Delta}$ satisfy $\tilde{\omega}=\omega+\frac{1}{2\tau}\frac{\tilde{\omega}}{\sqrt{\tilde{\Delta}^{2}-\tilde{\omega}^{2}}}$
and $\tilde{\Delta}=\Delta+\frac{1}{2\tau}\frac{\tilde{\Delta}}{\sqrt{\tilde{\Delta}^{2}-\tilde{\omega}^{2}}}$.
The self-energy $\Sigma_{w}$ therefore gives a superconducting gap
of size $\Gamma\left(\ll\Delta\right)$ on the wire. Since ${\rm Im}\Sigma_{w}\left(\omega=0\right)=0$,
we see that disorder is \emph{not} induced on the wire in this limit.
This is the gist of the weak-coupling result obtained earlier by Lutchyn
\emph{et~al}. \cite{Lutchyn2012} establishing the immunity of the
proximity-induced topological superconductivity to any disorder in
the SC itself, and it is only valid for $\Gamma\ll\Delta$ when the
induced gap is extremely small ($\sim\Gamma$).

In the strong-coupling limit in which the second term of Eq.~(\ref{eq:GS})
is not small and cannot therefore be ignored, both superconducting
gap and disorder are induced on the wire by the disordered SC. We
now investigate their dependence on the strength of the bulk disorder
and of the SM-SC coupling.

\subsection{Superconducting Gap Induced on the Wire\label{sub:inducedgap}}

We first investigate the superconducting gap induced on the wire which,
in the absence of Zeeman term on the wire, is equal to the induced
pairing potential $\Delta_{w}$. For simplicity we set $B=\alpha=0$
and $\mu=\Delta$ on the wire. (We have explicitly numerically checked
that our results presented here are generic, and using different parameter
values do not change the results at all qualitatively.) We note that
the natural energy scale of the problem is $\Delta$ and the natural
length scale is $\sqrt{k_{F}^{-1}\xi}$. The dimensionless parameter
quantifying disorder is defined as $d=\sqrt{k_{F}^{-1}\xi}/l=\sqrt{\frac{\xi}{l}\frac{1}{k_{F}l}}$,
where $l$ is the transport mean free path of the bulk SC.

\begin{figure}
\begin{centering}
\includegraphics[width=0.8\columnwidth]{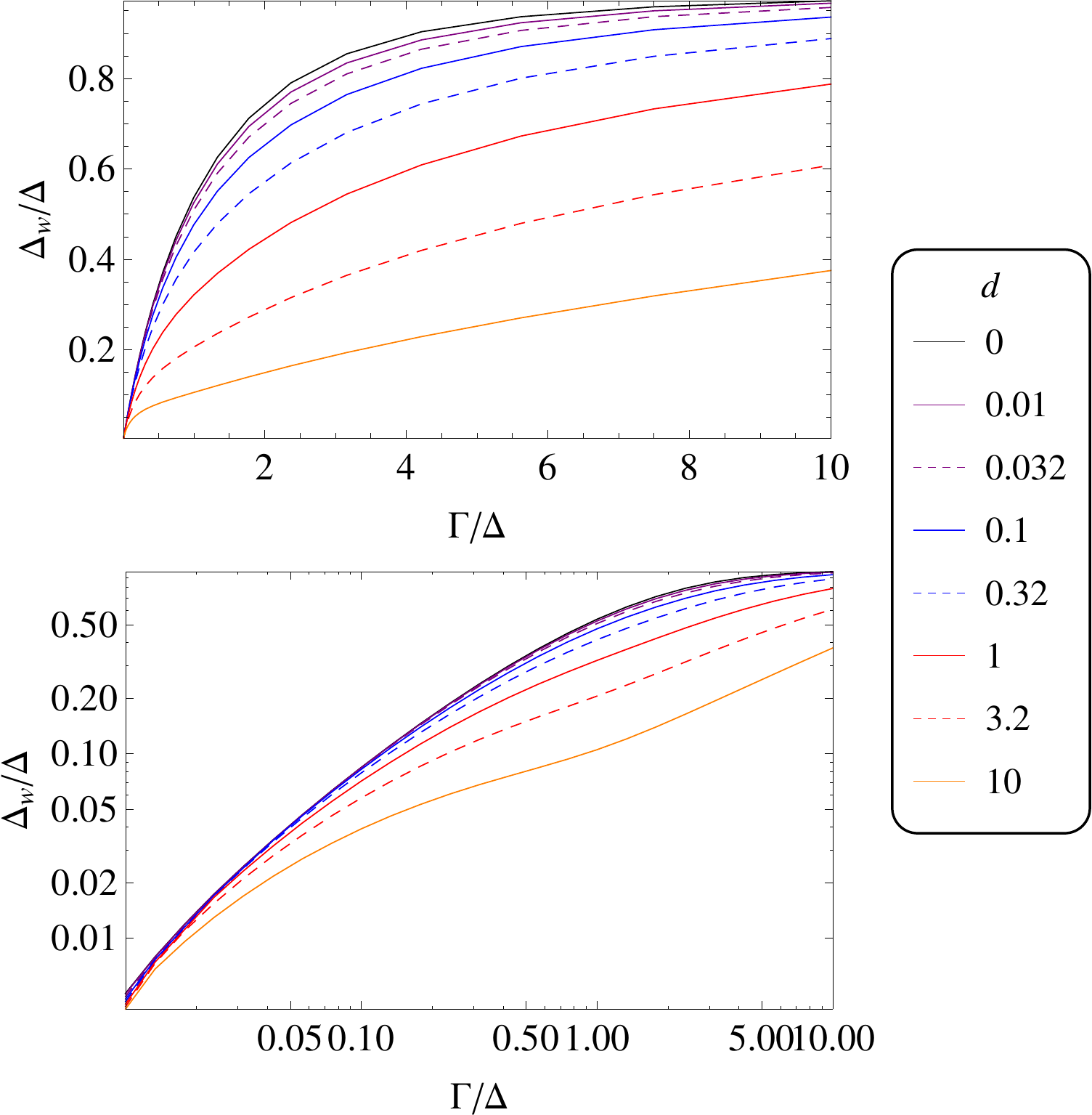}
\par\end{centering}

\caption{(Color online) Induced superconducting gap on the wire $\left(\Delta_{w}\right)$
against SM-SC tunnel coupling $\left(\Gamma\right)$ in (a) linear
scale and (b) logarithmic scale. Different lines correspond to different
strengths of disorder in the bulk of the SC $\left(d=\sqrt{k_{F}^{-1}\xi}/l\right)$.
The parameters on the wire are chosen to be $B=\alpha=0$ and $\mu=\Delta$.
\label{fig:GapGamma}}
\end{figure}

\begin{figure}
\begin{centering}
\includegraphics[width=1\columnwidth]{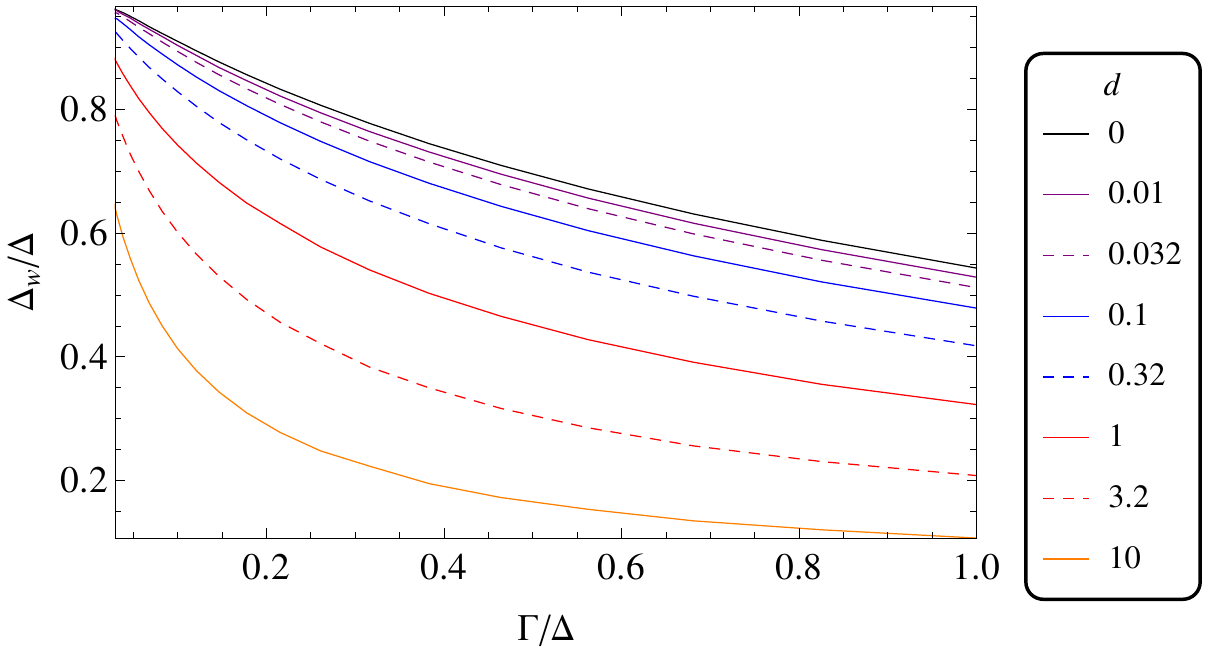}
\par\end{centering}

\caption{(Color online) Calculated dimensionless induced gap $\left(\Delta_{w}/\Gamma\right)$
plotted against the dimensionless tunnel coupling $\left(\Delta_{w}/\Gamma\right)$
for various values of the disorder parametrized by $d=\sqrt{k_{F}^{-1}\xi}/l$
as shown. The parameters on the wire are chosen to be $B=\alpha=0$
and $\mu=\Delta$.\label{fig:GapoverGamma}}
\end{figure}

In Fig.~\ref{fig:GapGamma}, we plot the calculated induced gap $\Delta_{w}$
as a function of $\Gamma$ in both linear and logarithmic scales.
We observe from the log-log plot that in the weak-coupling and weak
disorder limit, the induced gap scales linearly with coupling, i.e.,
\begin{equation}
\Delta_{w}\propto\Gamma,\;\mbox{for }l\gg\sqrt{k_{F}^{-1}\xi}\:\mbox{and}\:\Gamma\ll\Delta.\label{eq:Delta_wire_scaling}
\end{equation}
This is the same as previous calculations done with a clean SC \cite{Sau2010c}.
Perturbative calculations \cite{Lutchyn2012} have also shown that
at weak enough coupling $\Gamma$, disorder in the bulk SC does not
change the linear scaling of $\Delta_{w}$ with $\Gamma$. To verify
this, in Fig.~\ref{fig:GapoverGamma} we plot $\Delta_{w}/\Gamma$
against $\Gamma$, where we observe that
\begin{equation}
\lim_{\Gamma\rightarrow0}\frac{\Delta_{w}}{\Gamma}=1,\label{eq:Delta_wire_scaling2}
\end{equation}
irrespective of the strength of disorder in the SC, which is identical
to the induced gap for a \emph{clean} SC in the weak-coupling limit
$\left(\Gamma\ll\Delta\right)$, as we pointed out in the discussion
following Eq.~(\ref{eq:cleanInducedGap}). This is an indication
that the induced pairing dominates over the induced disorder in the
weak-coupling limit $\Gamma\ll\Delta$. In the following section,
we develop a more quantitative understanding by computing the scaling
behavior of disorder with respect to $\Gamma$ and comparing it with
Eq.~(\ref{eq:Delta_wire_scaling}), explicitly demonstrating that
the induced pairing dominates at small $\Gamma$.

\subsection{Disorder Induced on the Wire\label{sub:induceddis}}

Apart from the proximity-induced superconducting gap, the wire also
inherits disorder from the SC. In the topological phase, the superconducting
gap protecting the Majorana fermions could be destroyed by the induced
disorder if its strength becomes comparable to the gap \cite{Sau2012a}.
It is therefore important to study the dependence of the induced disorder
on the coupling strength so as to compare its strength with that of
the gap. The disorder on the wire has been defined in Eq.~(\ref{eq:disorderDef}),
which when written out in full is 
\begin{equation}
\tau_{w}^{-1}=\frac{{\rm TrIm}\Sigma_{w}\left(\omega=0\right)}{{\rm TrRe}\Sigma_{w}'\left(\omega=0\right)-4}.
\end{equation}
We remark here that $\tau_{w}^{-1}$ could, in general, be defined
as a frequency-dependent quantity, but for our purpose of investigating
its dependence on coupling strength we take only its value at zero
frequency. Also, from Eqs.~(\ref{eqs:GwireSC}), we see that ${\rm Im}\Sigma_{w}\left(\omega\right)$
is nonzero only if the band dispersion of the wire crosses the energy
$\omega$. Therefore, in order to produce a nonzero $\tau_{w}^{-1}$,
we choose the Zeeman term on the wire to be $B=5\Delta$, while keeping
$\alpha=0$ and $\mu=\Delta$, noting that the results depend only
weakly on the parameters.

\begin{figure}
\begin{centering}
\includegraphics[width=0.8\columnwidth]{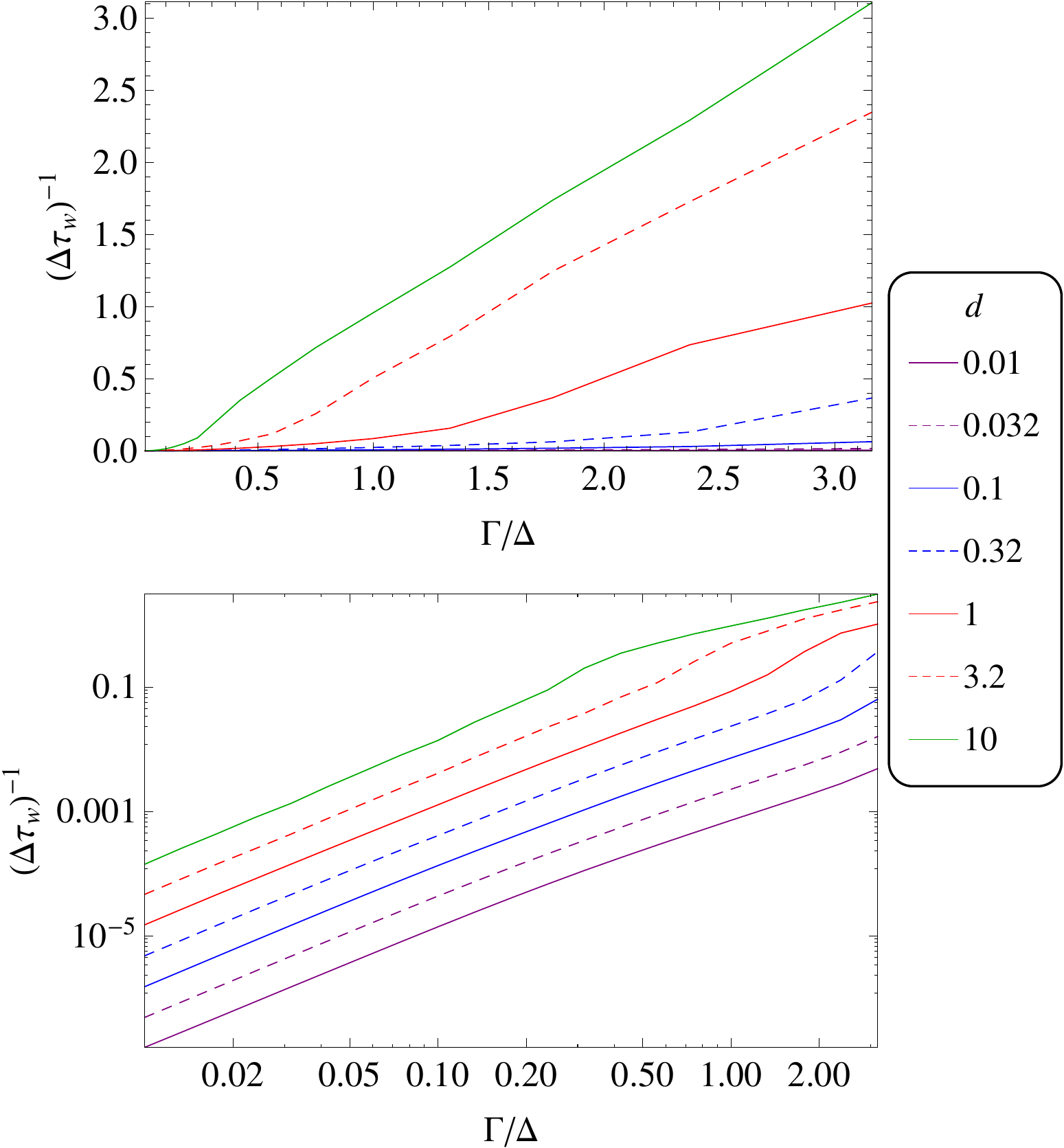}
\par\end{centering}

\caption{(Color online) Effective disorder strength on the wire $\left(\tau_{{\rm w}}^{-1}\right)$
as a function of SM-SC coupling $\left(\Gamma\right)$ in (a) linear
scale and (b) logarithmic scale. The parameters on the wire are chosen
as $\alpha=0$, $\mu=\Delta$, and $B=5\Delta$. Different lines represent
different disorder strength defined by $d=\sqrt{k_{F}^{-1}\xi}/l$.\label{fig:sigImGamma}}
\end{figure}

\begin{figure}
\begin{centering}
\includegraphics[width=0.8\columnwidth]{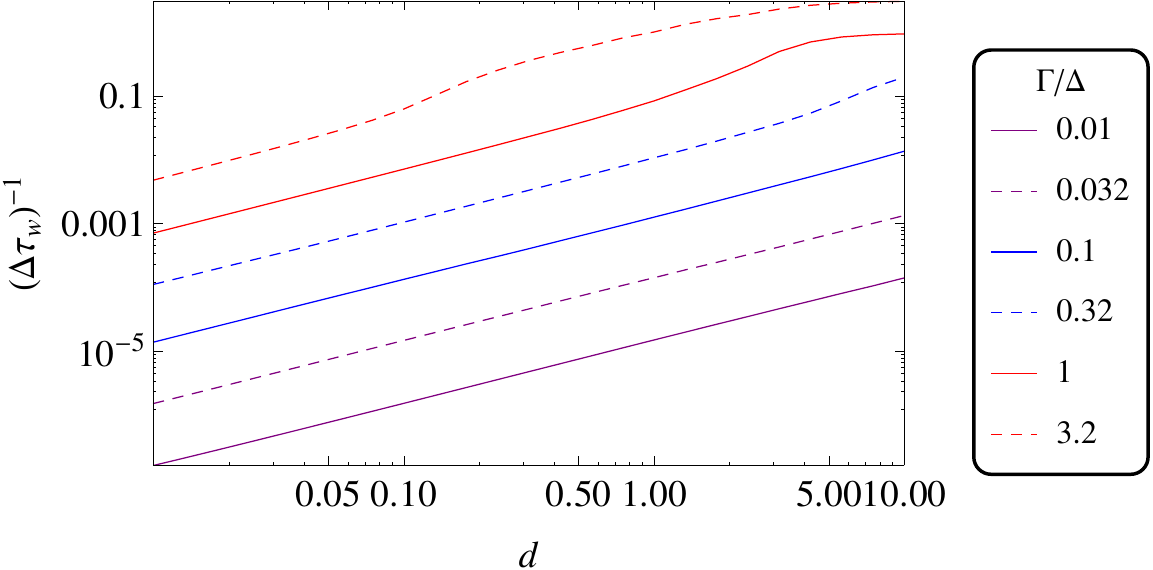}
\par\end{centering}

\caption{(Color online) Effective disorder on the wire $\left(\tau_{{\rm w}}^{-1}\right)$
against disorder in bulk SC $\left(l^{-1}\right)$ in logarithmic
scale. The parameters on the wire are chosen as $\alpha=0$, $\mu=\Delta$,
and $B=5\Delta$. The disorder strengths are defined by $d=\sqrt{k_{F}^{-1}\xi}/l$.\label{fig:sigImDis}}
\end{figure}

In Fig.~\ref{fig:sigImGamma}, we plot $\tau_{{\rm w}}^{-1}$ against
$\Gamma$ in linear and logarithmic scales. It shows that $\tau_{{\rm w}}^{-1}$
scales quadratically with $\Gamma$ when $\tau_{{\rm w}}^{-1}\ll\Gamma$.
However when $\tau_{{\rm w}}^{-1}$ is comparable with $\Gamma$,
the dependence changes to linear. Similarly, from the plot of $\tau_{{\rm w}}^{-1}$
against $l^{-1}$ in Fig.~\ref{fig:sigImDis}, we see that it scales
linearly when $l^{-1}$ is small. In summary, we have 
\begin{equation}
\tau_{w}^{-1}\propto\begin{cases}
l^{-1}\Gamma^{2}, & \tau_{{\rm w}}^{-1}\ll\Gamma,\\
\Gamma, & \tau_{{\rm w}}^{-1}\lesssim\Gamma.
\end{cases}\label{eq:Disorder_on_wire_scaling}
\end{equation}

This result could be compared with previous results obtained from
perturbative treatments which apply in the limit of very small induced
disorder \cite{Potter2011,Lutchyn2012}. We note that in addition
to recovering the quadratic scaling at weak coupling, our result also
shows a crossover to linear scaling at intermediate coupling strength,
which cannot be captured by perturbative approaches. This qualitatively
different linear scaling behavior of the induced disorder has important
consequences for the Majorana-carrying SC-SM hybrid nanowire systems
as discussed below.

\subsection{Suppression of Topological Gap by Disorder}

For practical reasons, the calculations in Secs.~\ref{sub:induceddis}
and \ref{sub:inducedgap} have been performed for wires with and without
Zeeman terms, respectively. It is natural to ask whether there is
a system in which both induced SC gap and induced disorder are at
play. The Majorana nanowires, with nonzero SOC and Zeeman terms, are
such systems. If its normal-state dispersion crosses zero energy,
the disorder term as defined by Eq.~(\ref{eq:disorderDef}) is nonzero.
On the other hand, with the SOC term present, the induced pairing
can produce a spectral gap even in the presence of the Zeeman term.
Therefore, a comparison between the scaling behaviors of induced pairing
and induced disorder with respect to the SC-SM tunnel coupling is
in order.

Comparing Eq.~(\ref{eq:Delta_wire_scaling}) and Eq.~(\ref{eq:Disorder_on_wire_scaling}),
we see that the scaling exponent $\left(2\right)$ of disorder $\left(\tau_{w}^{-1}\right)$
with respect to coupling strength $\left(\Gamma\right)$ is larger
than that $\left(1\right)$ of the pairing term $\left(\Delta_{w}\right)$.
This implies that $\Delta_{w}$ dominates at smaller $\Gamma$, while
$\tau_{w}^{-1}$ dominates at larger $\Gamma$. Therefore, increasing
the coupling strength does not always lead to a larger proximity-induced
topological gap since the induced effective disorder increases faster.
Rather, the optimal value of $\Gamma$ at which the topological gap
is maximum is dependent on the parameters of the system, and one expects
some intermediate value of $\Gamma$, which is neither too small (so
that the intrinsic topological gap itself is not too small) nor too
large (so that the induced disorder is not too strong overwhelming
the induced gap), to be the optimal choice. Purely on dimensional
grounds, the optimal value of the tunnel coupling is expected to be
$\Gamma\sim\Delta$ in the relatively clean-SC limit, but with increasing
SC disorder, we expect the optimal value of $\Gamma$ to decrease
so as to keep induced disorder effects small.

\begin{figure}
\begin{centering}
\includegraphics[width=1\columnwidth]{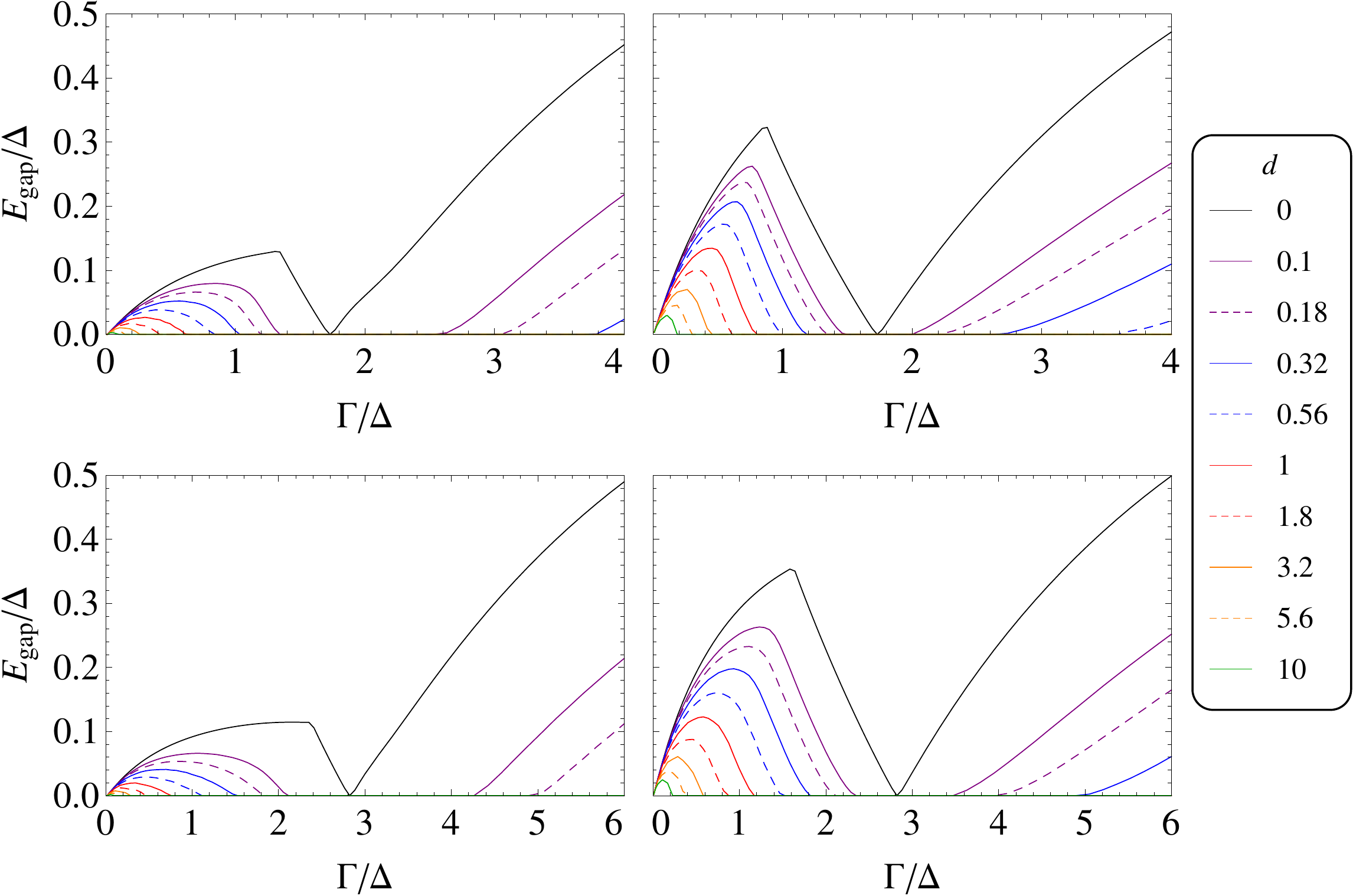}
\par\end{centering}

\caption{(Color online) Superconducting gap on the SM wire $\left(E_{{\rm gap}}\right)$
as a function of coupling strength $\left(\Gamma\right)$ for various
values of disorder strengths in the bulk SC $\left(d=\sqrt{k_{F}^{-1}\xi}/l\right)$.
The parameters on the wire are chosen to be $\mu=\Delta$ and (a)
$B=2\Delta$, $\alpha=0.3\sqrt{\frac{\Delta}{2m}}$, (b) $B=2\Delta$,
$\alpha=\sqrt{\frac{\Delta}{2m}}$, (c) $B=3\Delta$, $\alpha=0.3\sqrt{\frac{\Delta}{2m}}$,
(d) $B=3\Delta$, $\alpha=\sqrt{\frac{\Delta}{2m}}$.\label{fig:TopGap}}
\end{figure}

Figure~\ref{fig:TopGap} shows the superconducting gap on the wire
$\left(E_{{\rm gap}}\right)$ as a function $\Gamma$ for several
values of $B$ and $\alpha$. For $0<\Gamma<\sqrt{B^{2}-\mu^{2}}$,
the system is in a topological regime since the $s$-wave pairing
term is smaller than the Zeeman term \cite{Lutchyn2011,Oreg2010}.
When $\alpha$ is nonzero and in the absence of disorder (black line
in Fig.~\ref{fig:TopGap}), a topological gap exists and a zero-energy
Majorana fermion is present at each end of the nanowire. When $\Gamma$
increases to values near the Zeeman energy $\left|B\right|$, the
topological gap shrinks and closes completely at $\Gamma=\sqrt{B^{2}-\mu^{2}}$,
indicating the onset of the topological phase transition. The gap
then reopens at $\Gamma>\sqrt{B^{2}-\mu^{2}}$, where the system is
non-topological.

\begin{figure}
\begin{centering}
\includegraphics[width=0.8\columnwidth]{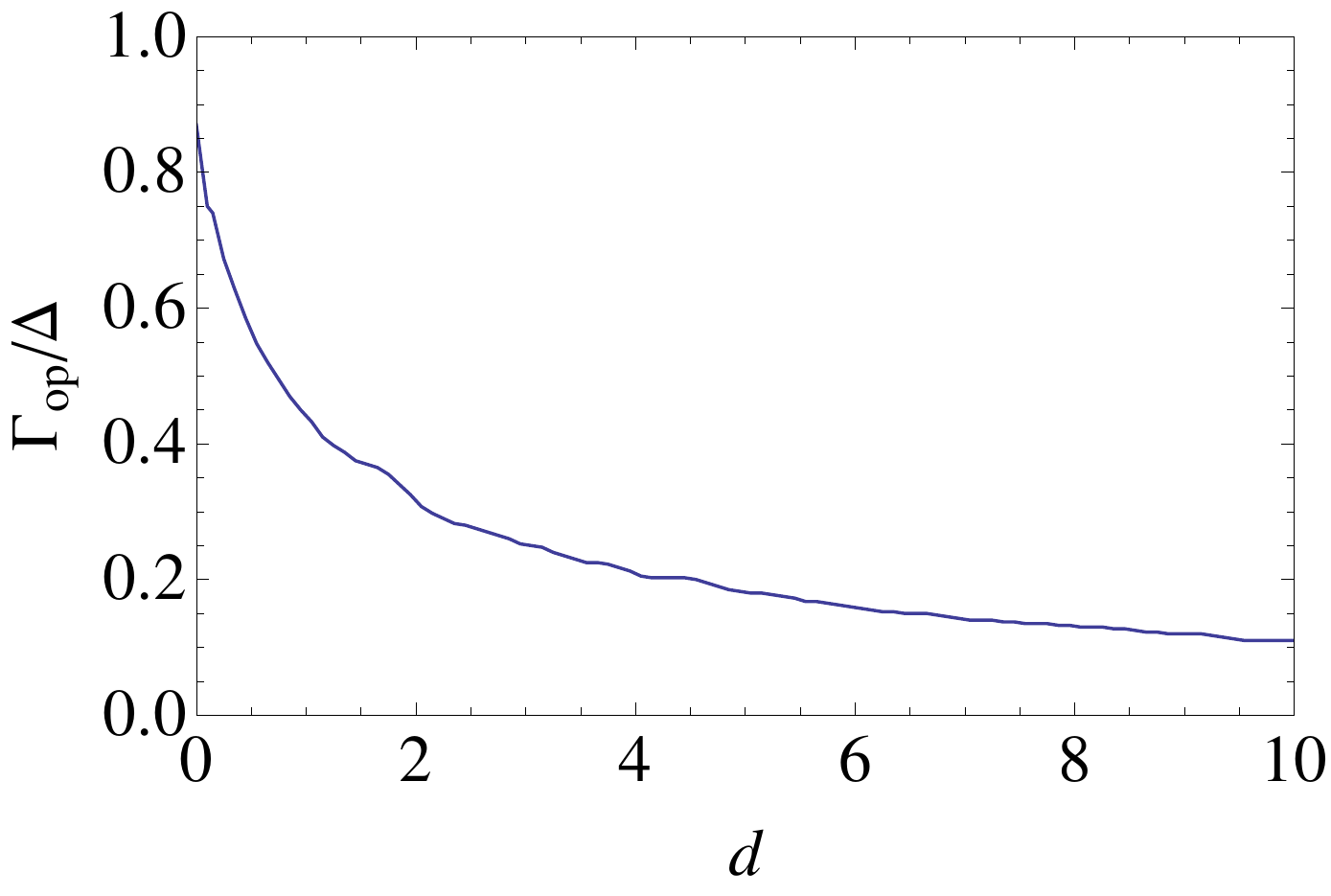}
\par\end{centering}

\caption{\label{fig:OpGamma}(Color online) The optimal value of $\Gamma$
at which the topological gap on the wire is maximum $\left(\Gamma_{{\rm op}}\right)$,
as a function of disorder in the bulk SC$\left(d=\sqrt{k_{F}^{-1}\xi}/l\right)$.
The parameters of the wire are equal to those of Fig.~\ref{fig:TopGap}(b),
i.e., $\mu=\Delta$, $B=2\Delta$, and $\alpha=\sqrt{\frac{\Delta}{2m}}$.}
\end{figure}

The gap on the wire is suppressed by disorder in the bulk SC, but
the degree of this suppression is dependent on $\Gamma$. In particular,
we see from Fig.~\ref{fig:TopGap} that the effect of disorder vanishes
when $\Gamma\rightarrow0$, and a topological gap (albeit small) exists
in that limit. This is consistent with our results in Secs.~\ref{sub:inducedgap}
and \ref{sub:induceddis} on the scaling of $\Delta_{w}$ and $\tau_{w}^{-1}$
with respect to $\Gamma$ {[}Eq.~(\ref{eq:Delta_wire_scaling}) and
Eq.~(\ref{eq:Disorder_on_wire_scaling}), respectively{]}: since
at small $\Gamma$, $\Delta_{w}$ scales as $\Gamma$ while $\tau_{w}^{-1}$
scales as $\Gamma^{2}$, the pairing term eventually dominates over
disorder at small $\Gamma$, producing a superconducting gap.

Figure~\ref{fig:OpGamma} shows the dependence of the optimal values
of tunnel coupling for which the topological gap on the wire is largest,
as a function of the disorder in the bulk SC. We see that with a more
disordered SC, it is actually more favorable to have a \emph{smaller}
coupling between the SM and the SC in order to generate a larger topological
gap. We believe this nonintuitive finding (Fig.~\ref{fig:OpGamma})
to be an important result for the fabrication of optimal SC-SM hybrid
structures for the realization of Majorana fermions--- the tunnel
coupling could be strong for an ultraclean SC (where $l\gg\xi$),
but for dirty SCs, one is far better off (as shown in Fig.~\ref{fig:OpGamma})
having a rather small SC-SM tunnel coupling.

\subsection{Discussion}

We have analyzed, within the framework of self-consistent Born approximation,
the effect of disorder residing solely in the bulk of the SC on the
spectral properties of the proximity-induced topological superconductivity
in the SC-SM hybrid system. The dependence of the induced pairing
gap and the induced disorder on the coupling strength is theoretically
explored. Crucially, we find that the topological gap induced on a
SM wire with SOC and Zeeman splitting can be very susceptible to the
disorder in the bulk SC when the SC-SM tunnel coupling is strong.
While the specific optimal coupling strength depends on the details
of the system, in general with high disorder in the SC a weak SM-SC
coupling is preferable.

These results have implications for the ongoing experimental efforts
to generate Majorana fermions by proximitizing a spin-orbit-coupled
SM nanowire under magnetic field in contact with a SC. Although it
is important to improve the interface quality between the two materials
so as to generate a hard gap on the SM \cite{Takei2013,Chang2015},
one should also be aware that a strong SC-SM tunnel coupling induces
stronger disorder on the SM wire, if the SC is diffusive. Thus it
is necessary to either use an ultraclean SC or to introduce a barrier
between the SM and the SC so as to effectively reduce the coupling
strength.

\section{Conclusion\label{sec:Conclusion}}

In this paper, we examined the effect of disorder in the bulk SC on
a Shiba state or a proximate SM in the context of the current search
for Majorana fermions in hybrid superconducting systems. In both cases,
we found that this type of disorder can have significant detrimental
impact, and could be an obstacle to create topological superconductivity
in the hybrid systems. In particular, disorder in the bulk SC can
randomly shift the energy of the Shiba states in the ferromagnet-superconductor
hybrid system, which is unfavorable for the existence of Majorana
modes since realizing a Majorana-carrying topological system requires
the fine tuning of the Shiba-state energy. (We mention that a complementary
model \cite{Dumitrescu2015,Hui2015} of the ferromagnetic adatom chain
on the superconductor system, which is adiabatically connected \cite{DasSarma2015,Peng2015}
to the Shiba model \cite{Brydon2015}, assumes the system to be equivalent
to the semiconductor nanowire on the superconductor structure, except
for the spin splitting in the adatom chain being extremely large so
that the system is completely spin polarized--- in such a spin-polarized
nanowire model of the ferromagnetic chain, the effect of bulk disorder
is qualitatively similar in the semiconductor and the ferromagnetic
chain system with disorder being detrimental in the strong-coupling
situation for reasons discussed in Sec.~\ref{sec:Nanowire} above.)
In the case of semiconductor-superconductor structure, the SM inherits
both superconducting pairing and disorder from the SC through the
proximity effect. We find that the scaling exponent of inherited disorder
with respect to the coupling strength between the two materials is
always larger than that of the inherited pairing. This implies that
while the pairing term can dominate over disorder and produce a spectral
gap on the SM at small coupling, upon increasing the coupling strength
the inherited disorder will eventually dominate and destroy the induced
SC gap. While the precise optimal value of the relevant coupling strength
for producing the strongest topological superconductivity depends
on the particular details of various parameter values, the key message
of our theory for the choice of the most suitable topological materials
parameters is that one should use ultraclean bulk superconductors
with extremely large normal-state low-temperature mean free path and
tune the tunnel coupling to a suitable value lower than the bulk superconducting
gap energy. We note that the disorder in the bulk superconductor enters
our theory through the dimensionless combination $d=\sqrt{\frac{1}{k_{F}l}\frac{\xi}{l}}$,
which implies that increasing either $k_{F}l$ or $l/\xi$ in the
parent superconductor should help to keep the disorder effects weak
in the system. This leads to our conclusion that among the commonly
used parent superconductors in the experimental hybrid systems, probably
Nb (Al) is the best (worst) choice, with Pb being somewhere in between
since typically Nb (Al) has the shortest (longest) coherence length,
thus making it easy (difficult) to satisfy the $l\gg\xi$ condition.
The detailed choice for the superconductor requires careful materials
preparation with the longest (shortest) possible values of the mean
free path (coherence length) in the system. The precise prediction
of our theory is simple: Choose a superconductor with the smallest
possible value of the dimensionless disorder parameter ``$d$'' and,
given this value of ``$d$'', tune the tunnel coupling so that it
equals $\Gamma_{{\rm op}}$, shown in our Fig.~\ref{fig:OpGamma}.
\begin{acknowledgments}
This work is supported by Microsoft Q, JQI-NSF-PFC, LPS-MPO-CMTC,
and the University of Maryland startup grant.
\end{acknowledgments}
\vfill{}

\bibliographystyle{apsrev4-1}
\bibliography{DisorderShiba}

\appendix
\end{document}